\documentclass{article}
\usepackage{graphicx} % Required for inserting images
\usepackage{amsmath}
\usepackage{cite}
\usepackage{neuralnetwork}
\usepackage{tikz}
\usetikzlibrary{positioning, calc}
\usepackage{algorithm}
\usepackage{algcompatible}
\usepackage{amsfonts}
\usepackage{float}
\usepackage{comment}
\usepackage{hyperref}
\usepackage{authblk}

\title{Neural Networks and Schramm-Loewner Evolutions}
\author[1]{Neilesh Shrotri} 
\author[2]{Vlad Margarint}
\affil[1]{Columbia University in the City of New York}
\affil[2]{University of North Carolina at Charlotte , Web: \href{https://margarintvlad.com/}{\text{margarintvlad.com}}.}
\date{June 2025}

\begin{document}

\maketitle

\begin{abstract}
In this manuscript, we explore the application of neural networks to predict the natural parameter $\kappa \geq 0$ of Schramm-Loewner Evolution (SLE$_\kappa$) theory. SLE$_\kappa$ is a family of random fractal curves that has significant implications in Statistical Mechanics and Conformal Field Theory. This parameter $\kappa \geq 0$ plays an important role in the theory as there are models of Planar Statistical Physics that are proven to have SLE as scaling limits as well as others that are conjectured to have this limit for various choices of the parameter $\kappa \geq 0$. In addition, there are three different statistical behaviors of the SLE curves as the parameter $\kappa$ changes in $[0, \infty).$ Leveraging the powerful pattern recognition capabilities of neural networks, this study aims to develop a predictive model that can estimate the $\kappa$ parameter with good accuracy.
\end{abstract}

\section{Introduction}
Schramm-Loewner Evolution (SLE or SLE$_\kappa$), introduced by Schramm in 2000 \cite{schramm2000}, describes a family of random fractal curves generated by solving the Loewner differential equation with a Brownian motion driver parameterized by a natural parameter $\kappa \geq 0$ . The SLE$_\kappa$ process has been pivotal in understanding critical phenomena in two-dimensional systems, such as Loop-Erased Random Walk for the choice $\kappa = 2$ \cite{k2}, Ising Model Interfaces (for $\kappa = 3$) \cite{k3}, and critical percolation on the triangular lattice ($\kappa = 6$) \cite{k6}. We refer the reader to \cite{lawler2005conformal}.
for a very detailed introduction to SLE theory. 

The parameter $\kappa \geq 0$ in SLE$_\kappa$ governs the behavior and geometric properties of the curves. Researchers studied various questions involving this parameter. For example, a natural question that was studied in the recent years was the continuity of the SLE curves and related objects in this parameter $\kappa$. We refer the reader to \cite{Fredcont}, \cite{Petercont}, \cite{Vladcont}, \cite{Vladweldcont}, \cite{VladJiaming} for studies concerning the continuity in the parameter $\kappa$ of the SLE curves and related objects.

In recent years, neural networks have emerged as a powerful tool for pattern recognition and prediction in various fields, including physics. Neural networks, particularly deep learning models, are capable of learning complex, nonlinear relationships from data, making them suitable for tasks such as predicting the $\kappa$ parameter in SLE$_\kappa$ theory. Previous studies have demonstrated the potential of machine learning in enhancing our understanding of physical systems \cite{carleo2019machine}.

This study aims to harness the capabilities of neural networks to predict the $\kappa$ parameter in SLE$_\kappa$ theory. By training a neural network model on simulated SLE$_\kappa$ data, we seek to develop a  predictive model. The findings of this research could contribute to a deeper understanding of critical phenomena and the broader application of machine learning techniques in theoretical physics. 

There are Planar Statistical Physics models which are conjectured to be described by the SLE in the scaling limit. However, though the models have an associated \(\kappa\) parameter, they have not been verified yet. We present a method that has promise in predicting the \(\kappa\) value given an SLE trajectory. In the future, we hope to predict the \(\kappa\) value from interfaces, with the end goal being to predict \(\kappa\) from a lattice model. 

As this is the first work, to the best of our knowledge, that is connecting Neural Networks with SLE theory, a very popular topic in Planar Statistical Physics models, we used some initial Neural Networks  models which were easier to train with enough computational power. We believe in the future one should try to investigate the connection between the two areas using more powerful Machine Learning models of Neural Networks.

Our study is structured as follows:
\begin{itemize}
    \item \textbf{Section 1: Introduction} - Provides a background on SLE$_\kappa$ theory and its significance in Statistical Mechanics and Conformal Field Theory. It discusses the importance of accurately predicting the $\kappa$ parameter and the potential of neural networks in this domain.
    \item \textbf{Section 2: Deterministic Loewner ODE} - Explains the Loewner Differential Equation (ODE,  for short) and its exact solution for a specific driving function. This section also describes the development of a neural network model to predict the driving function constant $c \geq 0$, including the model architecture, training process, and results. In our approach, we use 2-layer Neural Network along with the explicit solution to study the corresponding Loewner dynamics. As a methodology we train on $80 \%$ of the data and we test on $20\%$.
    \item \textbf{Section 3: Schramm-Loewner Evolution (SLE$_\kappa$)} - Introduces some elements of the Schramm-Loewner Evolution theory (SLE$_\kappa$) and presents the stochastic evolution describing the SLE$_\kappa$. It also covers the neural network model designed for predicting the $\kappa$ value in the SLE$_\kappa$ process, including data generation, model training, and evaluation of results.
     In our approach, we use 2-layer Neural Network along with an Euler-Maruyama Numerical Scheme to study the corresponding Stochastic Loewner dynamics. As before, as a methodology we train on $80 \%$ of the data and we test on $20\%$.
    \item \textbf{Sections 4 and 5} - Studies the application of Neural Networks to learn the parameter $\kappa\geq 0$ of the SLE traces. Compared with the previous sections, where we made neural networks learn the $\kappa\geq 0$ parameter from the stochastic Loewner dynamics inside the upper half-plane $\mathbb{H}=\{z \in \mathbb{C}| \text{Im} z >0\}$, the problem of learning the parameter of the traces is much more complicated as we need to evaluate the conformal maps on the boundary. For this, we use the Ninomiya-Victoir algorithm that we designed in a previous paper to deal with the challenge of simulating the SLE traces. See \cite{VladJamesTerry} for details. In addition, in these sections, we summarize the findings, evaluates the performance of the neural network models in predicting the $c \geq 0$ and $\kappa \geq 0$ terms, and suggests directions for future research.
\end{itemize}

\section{Deterministic Loewner Ordinary Differential Equation}

The Loewner differential equation, formulated by Charles Loewner in 1923, is a fundamental tool in complex analysis. It describes the evolution of conformal maps (that are transformations that preserve angles) and is expressed as:

\begin{equation}
\partial_t g_t(z) = \frac{2}{g_t(z) - \xi(t)}, \quad g_0(z) = z,
\end{equation}

where \(g_t(z)\) is a family of conformal maps and \(\xi(t)\) is a continuous real-valued function known as the driver function. In the deterministic case, the driver function is a specified function of time. A particular choice for \(\xi(t)\) can be for example \(\xi(t)=c\sqrt{t}\), which leads to an exact solution for the Loewner equation \cite{lawler2005conformal}. We refer the reader to \cite{tran2013convergencealgorithmsimulatingloewner} for other examples of driving functions for which the Loewner Differential Equation can be solved explictly.

\subsection{Exact Solution for \(\xi(t) = c\sqrt{t}\)}

For the driving function \(\xi(t) = c\sqrt{t}\), the exact solution of the Loewner differential equation can be described in terms of its inverse function. According to \cite{tran2013convergencealgorithmsimulatingloewner} the inverse of the Loewner map \(g_t(z)\) when \(\xi(t) = c\sqrt{t}\) is given by:

\begin{equation}
g_t^{-1}(z + c\sqrt{t}) = \left(z + 2\sqrt{t} \sqrt{\frac{a}{1-a}}\right)^{1-a} \left(z - 2\sqrt{t} \sqrt{\frac{a}{1-a}}\right)^{a},
\end{equation}

where \(a = \displaystyle \frac{1}{2} - \frac{1}{2} \frac{c}{\sqrt{16+c^2}}\).

This form of the solution provides insight into the behavior of the conformal map as it evolves over time for the given driving function.

\subsection{Implementation of the Exact Solution}

The implementation of the exact solution for \(\xi(t) = c\sqrt{t}\) can be demonstrated through computational methods available at the following link.\url{https://colab.research.google.com/gist/NShrotri/cc8d7cd81a13805e481df98735c823be/exactsol.ipynb} contains the code that computes and visualizes the solution.

We first present a pseudo-code version of the algorithm.

\begin{algorithm}[H]
\caption{Plot Inverse Loewner Map for \(\xi(t) = c\sqrt{t}\)}
\begin{algorithmic}[1]
\REQUIRE $c \gets 1.0,\; t_{\max} \gets 1.0,\; \mathit{num\_points} \gets 1000$
    \STATE $t \gets \mathrm{linspace}(0,\,t_{\max},\,\mathit{num\_points})$
    \STATE \textbf{function} InverseLoewnerMap($z,t,c$):
    \STATE\quad $a \gets 0.5 - 0.5\,\frac{c}{\sqrt{16 + c^2}}$
    \STATE\quad $\mathit{sqrt\_term} \gets 2\,\sqrt{t}\,\sqrt{\frac{a}{1 - a}}$
    \STATE\quad \textbf{return} \ $(z + \mathit{sqrt\_term})^{1 - a}\,(z - \mathit{sqrt\_term})^{a}$
    \STATE $\mathit{z\_values} \gets \mathrm{linspace}(-2,\,2,\,100)$
    \FOR{$z \in \mathit{z\_values}$}
      \STATE $g_{\mathrm{inv}} \gets \text{InverseLoewnerMap}(z,t,c)$
      \STATE \texttt{plot}$(t,\,g_{\mathrm{inv}})$\;\texttt{label=``z=\{z\}''}
    \ENDFOR
    \STATE \texttt{xlabel(``Time ($t$)'')}
    \STATE \texttt{ylabel(``$g_{t}^{-1}(z)$'')}
    \STATE \texttt{title(``Inverse Solution of Loewner ODE for $\xi(t)=c\sqrt{t}$'')}
    \STATE \texttt{legend()}
\end{algorithmic}
\end{algorithm}

\begin{comment}
For convenience of the reader, we also present a part of the code. 
\begin{verbatim}
import numpy as np
import matplotlib.pyplot as plt

# Parameters
c = 1.0  # Driving function constant
t_max = 1.0  # Maximum time
num_points = 1000  # Number of points in time

# Time array
t = np.linspace(0, t_max, num_points)

# Function to compute the inverse of the Loewner map
def inverse_loewner_map(z, t, c):
    a = 0.5 - 0.5 * c / np.sqrt(16 + c**2)
    sqrt_term = 2 * np.sqrt(t) * np.sqrt(a / (1 - a))
    return (z + sqrt_term)**(1 - a) * (z - sqrt_term)**a

# Initial z values (can be varied)
z_values = np.linspace(-2, 2, 100)

# Plotting the inverse solution
plt.figure(figsize=(10, 6))
for z in z_values:
    g_inv = inverse_loewner_map(z, t, c)
    plt.plot(t, g_inv, label=f'z={z:.1f}')
plt.xlabel('Time (t)')
plt.ylabel('g_t^{-1}(z)')
plt.title('Inverse Solution of Loewner ODE for $\xi(t) = c\sqrt{t}$')
plt.legend()
plt.show()
\end{verbatim}
\end{comment}

In this code, the inverse of the Loewner map is calculated for a range of initial points \(z \in \mathbb{H}=\{z \in \mathbb{C}| \text{Im}z\geq 0\}\) over time. The plot generated by this code illustrates how the conformal map evolves for different initial conditions. This implementation serves as a useful tool for visualizing and understanding the deterministic Loewner ODE with the specified driving function.

\subsection{Deterministic Neural Network Model}

To predict the constant \(c \geq 0\) appearing in the driving function \(\xi(t) = c\sqrt{t}\) in the deterministic Loewner ODE, we employed a neural network model using the Keras library in Python. The model architecture is summarized as follows:

\begin{verbatim}
model = Sequential([
    Dense(64, activation='relu', input_shape=(features.shape[1],)),
    Dense(64, activation='relu'),
    Dense(1)
])
\end{verbatim}

The Rectified Linear Unit (ReLU) activation used in our dense layers is defined as
\[
\mathrm{ReLU}(x) \;=\;\max\{0,\,x\},
\]
which zeroes out negative inputs while keeping positive inputs unchanged, improving computational efficiency and mitigating the vanishing‐gradient problem.

The input to the model is the data generated from the deterministic Loewner ODE, and the output is the predicted \(c\) term. \(c\) was sampled from \([0,3]\) from a uniform distribution. From the exact solution, we generated 10,000 trajectories as data points and had 80\% of the trajectories as training data and the remaining 20\% as testing data.

\subsection{Trajectory Computation via Inversion of the Exact Inverse Solution}

In this section, we describe our implementation method and strategy. To obtain the forward Loewner map \(g_{t}(z)\) for random driving‐function parameters, we exploit the known closed‐form expression for its inverse 
\[
g_{t}^{-1}(w)
\;=\;
\bigl(w \;+\;\alpha\,\sqrt{t}\bigr)^{\,1 - a}\,
\bigl(w \;-\;\alpha\,\sqrt{t}\bigr)^{\,a},
\]
where 
\[
a \;=\; \frac12 \;-\; \frac{c}{2\,\sqrt{\,16 + c^{2}\,}}, 
\qquad
\alpha \;=\; 2\,\sqrt{\frac{a}{\,1 - a\,}}, 
\]
and \(c\in[0,3]\) is the driving‐function constant.  In particular, for each fixed \((c,t)\), the map $g_{t}^{-1}(z)$ is given exactly by the formula above. We compute \(g_{t}(z_{0})\) for a given starting point \(z_{0}\in\mathbb{H}\) by numerically inverting the equation
\[
g_{t}^{-1}(w) \;=\; z_{0}.
\]
Concretely, the following procedure is used:

First, select \(c\) uniformly at random in the interval \([0,3]\).  Independently, choose the initial point 
\[
z_{0} \;=\; x_{0} \;+\; i\,y_{0},
\quad
x_{0},\,y_{0} \;\sim\;\mathcal{U}(0,10).
\]
Fix a target time \(T > 0\) and a time‐step \(\Delta t > 0\).  Let \(N = \lfloor T / \Delta t\rfloor\) and define the time grid \(t_{n} = n\,\Delta t\) for \(n=0,1,\dots,N\).  For each \(n\), we wish to find the unique complex \(w_{n}\) satisfying
\[
g_{\,t_{n}}^{-1}(w_{n}) \;=\; z_{0}.
\]
Equivalently, one solves the nonlinear equation
\[
F_{n}(w) \;:=\; 
\bigl(w + \alpha\,\sqrt{t_{n}}\bigr)^{\,1 - a}\,
\bigl(w - \alpha\,\sqrt{t_{n}}\bigr)^{\,a}
\;-\; z_{0}
\;=\;0,
\]
where \(a,\,\alpha\) depend on the chosen \(c\).  Because \(F_{n}(w)\) is analytic in \(w\) away from the branch cuts \(\pm\alpha\,\sqrt{t_{n}}\), each root \(w_{n} = g_{\,t_{n}}(z_{0})\) can be located robustly using a high‐precision Newton–Raphson iteration.

In our implementation, we employ arbitrary‐precision arithmetic (e.g., via the {\tt mpmath} library) to guarantee that each inversion step achieves a prescribed tolerance (for example, \(\lvert F_{n}(w_{n})\rvert < 10^{-30}\)).  At the \(0\)-th step (\(n=0\)), note that \(g_{0}(z_{0})=z_{0}\) since \(\sqrt{t_{0}}=0\).  For \(n\ge1\), given a previous approximation \(w_{n-1}\approx g_{\,t_{n-1}}(z_{0})\), we initialize the Newton iteration at \(w = w_{n-1}\) and update
\[
w \;\mapsto\;
w \;-\; \frac{F_{n}(w)}{\,F_{n}'(w)\,},
\]
where
\[
F_{n}'(w)
\;=\;
(1 - a)\,\bigl(w + \alpha\,\sqrt{t_{n}}\bigr)^{-a}\,\bigl(w - \alpha\,\sqrt{t_{n}}\bigr)^{\,a}
\;+\;
a\,\bigl(w + \alpha\,\sqrt{t_{n}}\bigr)^{\,1 - a}\,\bigl(w - \alpha\,\sqrt{t_{n}}\bigr)^{\,a - 1}.
\]
We iterate until \(\lvert F_{n}(w)\rvert\) falls below the target tolerance at time \(t_{n}\).  Starting from \(w_{0}=z_{0}\), this produces a sequence \(\{w_{n}\}_{n=0}^{N}\), which by construction satisfies \(w_{n} = g_{\,t_{n}}(z_{0})\) to arbitrary precision.

Thus, for each draw of \(c\) and \(z_{0}\), we obtain the forward Loewner trajectory
\[
t \;\longmapsto\; g_{t}(z_{0}).
\]
Repeating this procedure for many independent realizations of \(c\in[0,3]\) and \(z_{0}\in\{x+iy: x,y\in[0,10]\}\) yields a Monte Carlo ensemble of high‐precision solutions \(\{\,g_{t}^{(i)}(z_{0}^{(i)})\}_{i=1}^{M}\).  These trajectories can then be used to analyze statistical properties of the Loewner evolution under the square‐root driving function.

\subsubsection{Deterministic Neural Network Architecture}
In this study, we utilized a sequential feedforward neural network to predict a continuous target variable. The model architecture consists of three dense (fully connected) layers. The first dense layer contains 64 neurons and employs the Rectified Linear Unit (ReLU) activation function. This layer also defines the input shape, which matches the number of features in the input data.

The second dense layer also has 64 neurons and uses the ReLU activation function to introduce non-linearity, allowing the network to capture complex patterns in the data. The final layer is a single-neuron dense layer with a linear activation function, appropriate for regression tasks aimed at predicting a continuous value. This straightforward yet effective architecture enables the model to learn from the input features and produce accurate continuous predictions.

Loss was computed using Mean Squared Error. 

\subsection{Evaluation Metrics}

In our experiments, the neural network is trained to predict a scalar \(c\) value. We compile the model with mean squared error (MSE) as the loss function:
\begin{verbatim}
model.compile(optimizer='adam', loss='mse', metrics=['mae'])
\end{verbatim}
After training, we evaluate on a held‐out test set \(\{(x_i, c_i)\}_{i=1}^{N_{\mathrm{test}}}\), where \(c_i\) is the true constant factor in the driver \(c_i\sqrt{t}\) of the Loewner ODE. The input \(x_i\), is the Loewner trajectory taken and  given the exact formula inverse of a point with real and imaginary parts selected between 0 and 10 uniformly with start and end times, as well as number of steps user specified. For our purposes, we took the start time to be \(t = 0.1\), the end time to be \(t = 1.0\), and the number of steps to be \(100\). The model’s prediction is denoted \(\widehat{c}_i = f_{\theta}(x_i)\) where \(\theta\) represents the collection of all trainable parameters (weights and biases) of the neural network.  Because we specified \texttt{loss='mse'}, the test loss is exactly the mean squared error over all test‐set examples:
\[
\text{Test Loss} \;=\; \mathrm{MSE}\bigl(\{c_i, \widehat{c}_i\}\bigr)
\;=\; \frac{1}{N_{\mathrm{test}}} \sum_{i=1}^{N_{\mathrm{test}}} \bigl(c_i - \widehat{c}_i\bigr)^{2}.
\]
By squaring each prediction error \(\bigl(c_i - \widehat{c}_i\bigr)\), larger deviations in the predicted \(\hat{c}\) are penalized more heavily, and dividing by \(N_{\mathrm{test}}\) normalizes the loss so that it does not scale with the dataset size. A lower MSE on the test set indicates that the network’s predicted \(\hat{c}\) values are, on average, closer to the true values.

\subsubsection{Results}
We received promising results from the neural network.

\begin{figure}[h]
    \centering
    \includegraphics[width=0.65\linewidth]{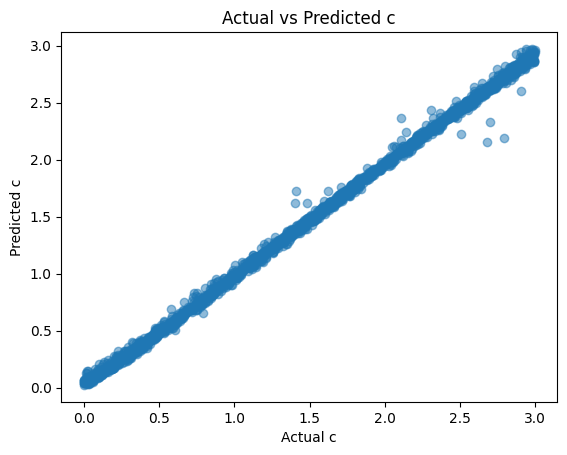}
\end{figure}

We achieved a test loss of 0.00264.

\section{Schramm-Loewner Evolutions}

\subsection{SLE\(_\kappa\) Differential Equation}

The SLE\(_\kappa\) dynamics is defined by a stochastic version of the Loewner differential equation 

\begin{equation}
\partial_t g_t(z) = \frac{2}{g_t(z) - \sqrt{\kappa} B_t}, \quad g_0(z) = z,
\end{equation}

where \(B_t\) is a standard one-dimensional real Brownian motion and \(\kappa \geq 0\) the aforementioned positive parameter that determines the properties of the curves.

The random nature of the real-valued Brownian motion \(B_t\) introduces significant complexity into the solutions of the Loewner differential equation. However, this randomness is also what enables Schramm Loewner Evolution to model a variety of critical phenomena in two-dimensional Statistical Physics, as we described in the introduction of our manuscript.

\subsection{Stochastic ODE Simulation Algorithm}

We wish to approximate, for many independent realizations, the solution \(g(t)\) of the stochastic ODE
\[
\frac{\mathrm{d}g(t)}{\mathrm{d}t}
\;=\;
\frac{2}{\,g(t)\;-\;\sqrt{\kappa}\,W(t)\,},
\qquad
g(0)=z,
\]
where \(W(t)\) is a standard one‐dimensional Brownian motion, \(\kappa>0\) is a parameter, and \(z\in \mathbb{H}\) is the initial condition.  To do this numerically, we fix a final time \(T>0\) and a time‐step \(\Delta t>0\), and set
\[
N \;=\; \left\lfloor\frac{T}{\Delta t}\right\rfloor
,\qquad
t_n = n\,\Delta t,\quad n=0,1,\dots,N.
\]
For each realization, the algorithm proceeds as follows:

First, draw \(\kappa\) from its prescribed distribution, and sample the initial point \(z\) in the complex plane.  We then approximate the Brownian motion \(\{W(t_n)\}_{n=0}^{N}\) by
\[
\Delta W_n \;\sim\;\mathcal{N}(0,\,\Delta t), 
\quad
W_n = \sum_{j=0}^{\,n-1\,} \Delta W_j,
\quad
n=0,1,\dots,N-1,
\]
with \(W_0 = 0\).  Next, we initialize the discrete solution array by setting
\[
g_0 \;=\; z.
\]
At each time step \(n=0,1,\dots,N-1\), we perform an explicit Euler update:
\[
g_{n+1}
\;=\;
g_n
\;+\;
\Delta t \;\frac{2}{\,g_n \;-\;\sqrt{\kappa}\,W_n\,}.
\]
Once we have computed \(\{g_n\}_{n=0}^{N}\) for that realization, we record both \(\kappa\) and the array \(\{g_n\}\).  Repeating this procedure for a total of \(M\) independent simulations yields three arrays of length \(M\): the sampled parameter values \(\{\kappa^{(i)}\}_{i=1}^M\), the randomly‐chosen initial points \(\{z^{(i)}\}_{i=1}^M\), and the corresponding numerical trajectories \(\{\,g^{(i)}_n\}_{i=1,\;n=0}^{M,\;N}\).

In summary, for each simulation index \(i=1,\dots,M\):  
\[
\begin{aligned}
&\text{(i) Sample } \kappa^{(i)},\quad z^{(i)},\\
&\text{(ii) Generate Brownian increments } \Delta W_n^{(i)} \sim \mathcal{N}(0,\Delta t),\quad W_n^{(i)} = \sum_{j=0}^{n-1} \Delta W_j^{(i)},\\
&\text{(iii) Set } g_0^{(i)} = z^{(i)}, \text{ and for } n=0,\dots,N-1\text{ compute}\\
&\quad g_{n+1}^{(i)} = g_n^{(i)} \;+\; \Delta t \;\frac{2}{\,g_n^{(i)} - \sqrt{\kappa^{(i)}}\,W_n^{(i)}\,},\\
&\text{(iv) Store } \kappa^{(i)},\;\{g_n^{(i)}\}_{n=0}^N\,.
\end{aligned}
\]
This method yields a Monte Carlo ensemble of discrete approximations to \(g(t)\) up to time \(T\).

\subsection{SLE\(_\kappa\) Neural Network Model}

To predict the \(\kappa\) term in the SLE\(_\kappa\) case, we modified the deterministic neural network model. The model architecture is summarized as follows:

\begin{algorithm}[H]
  \caption{Neural Network Architecture}\label{alg:nn-architecture}
  \begin{algorithmic}[1]
    \STATE \textbf{Input:} data of shape \(\texttt{input\_shape}\);
    \STATE \(x \gets \mathrm{Input}\bigl(\text{shape} = \texttt{input\_shape}\bigr)\);
    \STATE \(x \gets \mathrm{Flatten}(x)\);
    \STATE \(x \gets \mathrm{Dense}\bigl(x,\,128,\;\text{activation}=\text{ReLU}\bigr)\);
    \STATE \(x \gets \mathrm{Dropout}\bigl(x,\,\text{rate}=0.2\bigr)\);
    \STATE \(x \gets \mathrm{Dense}\bigl(x,\,64,\;\text{activation}=\text{ReLU}\bigr)\);
    \STATE \(x \gets \mathrm{Dropout}\bigl(x,\,\text{rate}=0.2\bigr)\);
    \STATE \(y \gets \mathrm{Dense}(x,\,1)\);
    \STATE \textbf{Output:} \(y\);
  \end{algorithmic}
\end{algorithm}

\begin{comment}
\begin{verbatim}
input_layer = Input(shape=input_shape)
flatten_layer = Flatten()(input_layer)
dense_layer_1 = Dense(128, activation='relu')(flatten_layer)
dropout_layer_1 = Dropout(0.2)(dense_layer_1)
dense_layer_2 = Dense(64, activation='relu')(dropout_layer_1)
dropout_layer_2 = Dropout(0.2)(dense_layer_2)
output_layer = Dense(1)(dropout_layer_2)
\end{verbatim}
\end{comment}

The input to the model is the data generated from SLE\(_\kappa\), and the output is the predicted \(\kappa\) value. We emphasize that the parameter \(\kappa\) was sampled from \([0,8]\) using a uniform distribution. We generated 10,000 trajectories as data points using the Euler-Maruyama method for the Loewner ODE dynamics and had 80\% of the trajectories as training data and the remaining 20\% as testing data.

\subsubsection{SLE\(_\kappa\) Neural Network Architecture}
Following the simpler sequential model, we also explored a more complex neural network architecture to enhance predictive performance. This architecture begins with an input layer tailored to the shape of the input data, followed by a flattening layer to convert the multi-dimensional input into a one-dimensional array. The first dense layer contains 128 neurons and utilizes the ReLU activation function to capture intricate patterns in the data. To address overfitting, a dropout layer with a rate of 0.2 is applied, randomly setting 20\% of the neurons to zero during training.

Subsequently, the data is processed by a second dense layer with 64 neurons, also employing the ReLU activation function. Another dropout layer with the same dropout rate follows to further reduce overfitting. The final output layer, consisting of a single neuron without an activation function outputs the predicted \(\kappa\) parameter. This more intricate architecture, with its additional dense layer and dropout mechanisms, aims to balance model complexity and generalizability, thereby enhancing the robustness of the predictions.

Like with the sequential model, loss was computed using Mean Squared Error.

\subsubsection{Results}
We considered in our analysis two scenarios. In the first scenario, we considered a fixed noise, that is a fixed Brownian motion trajectory as a driver. In the second case, we varied the noise by considering more samples paths of Brownoan motion driver. We received promising results from the neural network.

\begin{figure}[h]
    \centering
    \includegraphics[width=0.65\linewidth]{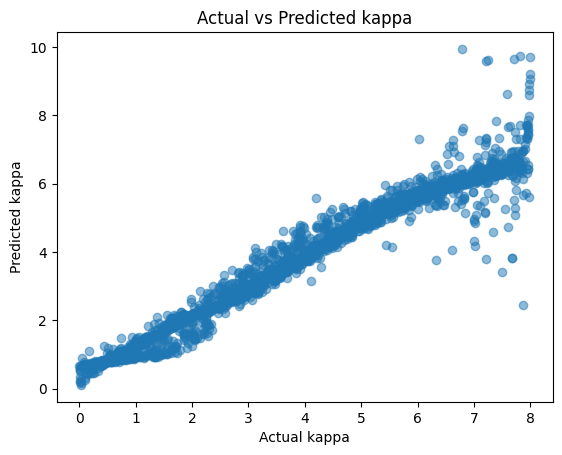}
    \caption{Same Noise}
\end{figure}

\begin{figure}[h!]
    \centering
    \includegraphics[width=0.65\linewidth]{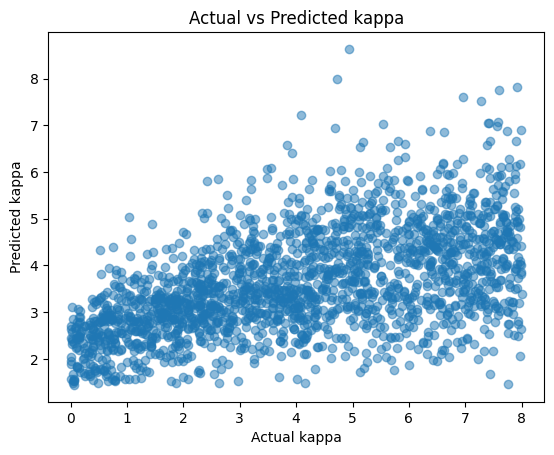}
    \caption{Different Noise}
\end{figure}

 \newpage

For the same noise case, we achieved a test loss of 0.345.

For the different noise case, we achieved a test loss of 3.98.

\section{Recent Developments: Predicting the $\kappa$ parameter of the SLE traces for a fixed Brownian Motion}

In the previous sections we made neural networks learn the parameter $c \geq 0$ in the deterministic case, or $\kappa \geq 0$ in the stochastic case from the Loewner trajectory, that is, we made them learn the conformal maps $g_t(z)$ that satisfy the Loewner ODE. In this part, we make the neural network learn the parameter $\kappa\geq 0$ corresponding to the SLE curves that are simulated with the  Ninomiya-Victoir (NV) algorithm (see \cite{VladJamesTerry}). This is a much harder problem as the SLE traces are defined as $\lim_{y \to 0+ }g_t^{-1}(iy)$, thus reaching the singularity of the Loewner ODE dynamics as we need to evaluate the conformal maps on the boundary.

Extending our previous exploration of neural networks applied to the Schramm-Loewner Evolution, we now investigate the scenario where the SLE trace is generated using a fixed Brownian motion. By fixing the noise (Brownian path) used in simulating the SLE trajectories, we aim to reduce variability in the input data and enhance the predictive capability of our neural network model.

The neural network architecture employed remains consistent with that described previously, consisting of an input layer, flattening layer, two dense hidden layers with 128 and 64 neurons respectively, and dropout layers to mitigate overfitting. The model is trained on data generated from the SLE$\kappa$ differential equation with a fixed Brownian motion path.

\begin{figure}[htbp]
    \centering
    \includegraphics[width=0.65\textwidth]{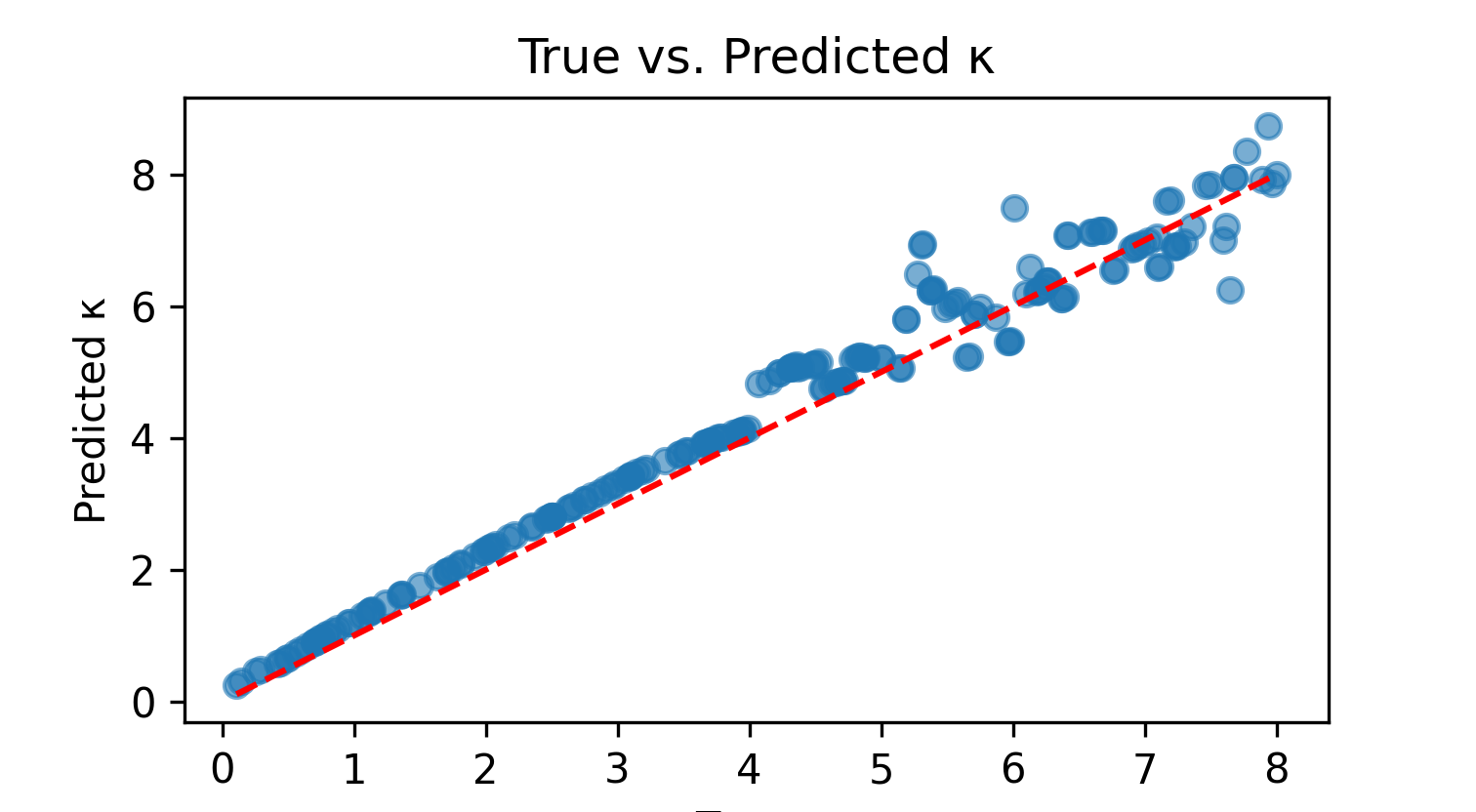}
    \caption{True vs.\ Predicted $\kappa$ with fixed Brownian motion. The dashed red line represents perfect prediction.}
    \label{fig:true_vs_predicted_kappa}
\end{figure}

Initial results for this fixed-noise scenario are highly encouraging. The model achieved a mean squared error (MSE) of $0.194$, indicating a significant improvement in prediction accuracy compared to the previous scenarios tested. Figure~\ref{fig:true_vs_predicted_kappa} illustrates the strong correlation between true and predicted values of $\kappa$, confirming the model's effectiveness under these conditions.

These promising outcomes suggest that neural networks can predict the \(\kappa\) parameter given an SLE trace. Future research directions will involve deeper exploration of the robustness and generalizability of the model under variable noise, as well as further optimization of neural network architecture to refine predictive accuracy.

Our experiments yielded promising results, demonstrating that neural networks can accurately predict the \(c\) and \(\kappa\) terms in the Schramm-Loewner Evolution (SLE). The first model, a sequential architecture with two hidden layers of 64 neurons each, using ReLU activations, effectively learned the patterns in the data generated by the deterministic Loewner ODE. This model achieved a test loss of 0.00264, indicating its efficiency in predicting the \(c\) parameter.

Building on this work, we explored a more complex neural network architecture for the SLE\(_\kappa\) case. This model included an input layer, a flattening layer, and two dense layers with 128 and 64 neurons respectively, both using ReLU activations, with dropout layers to mitigate overfitting. The final output layer consisted of a single output neuron predicting 
the \(\kappa\) parameter. This architecture balanced complexity and generalizability, achieving a test loss of 0.345 for the same noise case and 3.98 for different noise cases, highlighting the model's sensitivity to noise characteristics in the training data.

An important finding was the significant impact of \(T_{\text{end}}\), the stopping time of the simulation, on the model's accuracy. Larger \(T_{\text{end}}\) values led to better predictions, underscoring the importance of appropriate simulation parameters in training robust models.

\section{Future Directions}

As a first potential further direction of study we can consider the following. The machine learning model can be modified for example to a classification problem in the future. If one seeks to verify if a model is governed by an SLE with \(\kappa = \kappa'\), one can simulate 5,000 trajectories of SLE\(_{\kappa'}\) with the label 1, and 5,000 trajectories of SLE\(_{\kappa \neq \kappa'}\) with the label 0 and train the neural network with these new labels.

Secondly, our research has been constrained by memory limitations and computational power. Training more complex models and running simulations with higher \(T_{\text{end}}\) values require significant computational resources. These limitations affect our ability to explore larger datasets and more sophisticated architectures, potentially hindering the discovery of even more accurate and robust models.

In addition to better predicting the \(\kappa\) parameter, future research will focus on extending these neural network techniques to predict the \(\beta\) parameter of Dyson Brownian motion, which would serve as the driver function for the SLE. Dyson Brownian motion is a stochastic process with applications in random matrix theory and statistical mechanics. Accurately predicting the \(\beta\) parameter, which controls the interaction strength between particles, could provide valuable insights into the dynamics of complex systems and enhance the understanding of SLE driven by such processes.

To this end, optimizing the neural network architecture and exploring different regularization techniques will be crucial. Additionally, investigating the impact of various simulation parameters and noise characteristics on model performance will provide deeper insights into the robustness and applicability of neural networks for predicting parameters in both SLE and Dyson Brownian motion. Overcoming current computational limitations through access to more powerful hardware or more efficient algorithms will be essential for advancing this research.

\section{Acknowledgements}
This work was funded by the NSF-REU DMS-2150179 grant. We also thank Prof. Ivan Corwin for useful suggestions. 
\bibliographystyle{plain}
\bibliography{references}
\end{document}